\documentclass[prd,aps,twocolumn,superscriptaddress,preprintnumbers,floatfix,amssymb,amsmath]{revtex4}

\newcommand{\beqn}{\begin{eqnarray}}
\newcommand{\eeqn}{\end{eqnarray}}
\newcommand{\eq}[1]{(\ref{#1})}

\begin{document}

\preprint{ITEP-LAT/2009-04}

\title{Magnetic component of gluon plasma and its viscosity}

\author{M.N.~Chernodub}
\affiliation{LMPT, CNRS UMR 6083, F\'ed\'eration Denis Poisson,
Universit\'e de Tours, F-37200, Tours, France} \affiliation{DMPA,
University of Gent, Krijgslaan 281, S9, B-9000 Gent, Belgium}
\affiliation{ITEP, B.~Cheremushkinskaya 25, Moscow, 117218, Russia}
\author{H.~Verschelde}
\affiliation{DMPA, University of Gent, Krijgslaan 281, S9, B-9000 Gent, Belgium}
\author{V.I.~Zakharov}
\affiliation{ITEP, B.~Cheremushkinskaya 25, Moscow, 117218, Russia}
\affiliation{Max-Planck-Institut f\"ur Physik, F\"ohringer Ring 6, 80805 \"Munich, Germany}
\begin{abstract}
We discuss the role of the magnetic degrees of freedom of the gluon plasma in its viscosity.
The main assumption is that motions of the magnetic component and of the rest
of the plasma can be considered as independent. The magnetic component
in the deconfined phase is described by a three-dimensional (Euclidean)
field theory. The parameters of the theory can be estimated phenomenologically,
from the lattice data. It is not ruled out that the magnetic component is superfluid.
\end{abstract}

\maketitle

\section{Introduction}

The interpretation of heavy ion experiments at RHIC and first-principle lattice
simulations suggest that the quark-gluon plasma has quite unusual
properties~\cite{review,teaney}. Contrary to general expectations, at temperatures
just above the critical temperature, $T_c$, the plasma has properties rather of an
ideal fluid  than of a weakly interacting gas of quarks and gluons~\cite{foot1}.

The unexpectedly low  viscosity is in contrast with thermodynamical properties
of the plasma which do not betray much unexpected. Indeed,
at high enough temperatures
the difference between
the observed pressure (energy) density and its perturbative value
can be fitted by a $g^6(T)$ contribution,
\begin{equation}\label{fit}
{p(T)_{\mathrm{full}}-p(T)_{\mathrm{pert}}\over T^4}~\approx~ {\mathrm{const}} \cdot g^6(T)~~,
\end{equation}
where $g^2(T)$ s the running constant. The fit (\ref{fit}) is expected
on theoretical grounds since in the order $g^6(T)$ one runs into infrared divergencies
which can in fact be treated only non-perturbatively.
Thus the constant in the r.h.s. of Eq.~(\ref{fit}) is not calculable
analytically at this time.

Let us consider an idealized picture and assume for the moment that
the deviations of the thermodynamical properties from (weak-coupling) perturbative
values are small while the viscosity is much lower than its perturbative
value. This would suggest that we deal with two (quantum mechanically) independent
motions. Indeed if there are two independent fluid components with viscosities
$\eta_{1,2}$ then we have additivity of fluidity,
i.e., inverse viscosity~\cite{fluidity}:
\begin{equation}\label{independent}
{1\over \eta_{\mathrm{tot}}}~=~{(phase~space)_1\over \eta_1}~+~{(phase~space)_2\over \eta_2}~~.
\end{equation}
If, say, $\eta_2\approx0$ the total value $\eta_{\mathrm{tot}}$ can still be small even if
the corresponding phase space factor $(phase~space)_2$ is small.

In this note we will explore the possibility that the  magnetic component of the
Yang-Mills plasma \cite{magncom,shuryak,alessio,atsushi,sasha,ratti}.
provides us with an independent motion
in the sense of viscosity, see Eq. (\ref{independent}). Examples of ``independent motions''
in condensed-matter systems
are well known. In the case of ordinary superconductivity,
 the  contribution of Cooper pairs is independent of electrons in the normal state.
Closer to our problem, superfluid and ordinary (or dissipative)
components of liquids at low temperatures can be treated as independent~\cite{lifshitz}.

In a crude approximation, one can understand by magnetic component the 3d field theory which
determines the r.h.s. of Eq. (\ref{fit}) at high temperatures.
Indeed it is known since long \cite{pisarski} that at high temperatures it is the
3d field theory corresponding to the zero Matsubara frequency $\omega_M=0$
which is to be treated non-perturbatively. In this limit, the temperature dependencies
of all the non-perturbative observables can be reconstructed from their dimensions.
For example, the string tension of the spatial Wilson line is to be proportional
\begin{equation}\label{scaling}
\sigma_3~\sim~ g_3^4\,,
\end{equation}
where
\begin{equation}
\label{g3}
g^2_3 = g^2(T)T\,,
\end{equation}
is the dimensionally reduced gauge coupling, $g^2(T)$ is the running coupling of the 4d Yang-Mills theory.
Here $g^2(T)$ is assumed to be small enough to serve as a small expansion parameter.
Numerically, the scaling laws like (\ref{scaling}) set in at temperatures not too much
higher  than the critical temperature of the deconfining phase transition $T_c$
although $g^2(T)$ does not seem  yet to be small ($T_c/\Lambda_{QCD}\approx 1.2$ \cite{fingberg}).

More precisely,  the magnetic component is  defined in terms of the
magnetic monopoles and center vortices identified on the lattice.
These degrees of freedom are commonly believed to be responsible
for  confinement at low temperatures, for a review see, e.g.,
\cite{greensite}. As is argued in Refs.
\cite{magncom,shuryak,alessio,atsushi,sasha} at $T>T_c$ the magnetic degrees of freedom
become a part of the Yang-Mills plasma. A subtle point is that magnetic degrees of freedom are
studied on the lattices, or in Euclidean space while viscosity is defined most
straightforwardly in the Minkowski space. In particular, we
are going to treat the magnetic component as an ``independent motion'' in the Minkowski
space.

To substantiate (or reject) this hypothesis one needs a continuum-theory interpretation
of the lattice defects. Dual models of Yang-Mills theories, see in particular \cite{sasha} and
references therein, seem to provide such an interpretation. Namely, the dual models are formulated in
terms of strings living in extra dimensions, for a review see, e.g., \cite{aharony}. Then there exist
various topologically stable solutions in the dual formulation of the Yang-Mills theories.
In particular, the observed properties of the lattice vortices and monopoles fit
remarkably well the pattern expected within the dual models
for the magnetic strings \cite{zakharov,sasha}.
Moreover, the monopole and vortex pictures get unified since monopoles are to be thought
about as 1d defects (trajectories) living on the 2d defects (vortices, or strings)~\cite{zakharov,adriano}.
The monopoles and vortices constitute the magnetic component of the gluon plasma.

What is most relevant to our purposes, it is expected theoretically that
the magnetic strings become time oriented at $T>T_c$ \cite{sasha}
since  only time oriented magnetic strings are (nearly) tensionless
at $T>T_c$.
Then the magnetic strings reduce to their projections to a time-slice since the time
dependence is trivial. Thus, the solutions of the full 4d theory are mapped into
3d solutions. Consider now the 3d medium of these topological excitations.
Since we deal with solutions of the full theory we do not need
to consider further, for example, their interaction with gluons. The properties of the
3d medium  depend, however, on the interaction of the
topological excitations between themselves which is not taken into the account yet.
Similarly, in the theory of superconductivity one starts first with an (approximate) solution for
the Fermi-liquid at T=0. The Cooper pairs emerge after accounting for
(relatively weak) interactions near the edge of the Fermi-sphere.

At present, there are no means to clarify interaction among the solutions
theoretically and we will
rely on the lattice phenomenology at this point. The lattice data are in the
Euclidean space, however. In the static approximation for the magnetic defects
the continuation from the Euclidean to Minkowski space is trivial and this is the
approximation we will use. In the static approximation, the measurements reduce
to the measurements on the ground state of the 3d system (which is the same in
the Euclidean and Minkowski spaces). There is a spectrum of excitations which
determine, in particular the time development of the system.
The spectrum is obtained by quantization on the background of the classical solutions.
If there is time dependence, the continuation from the Euclidean to Minkowski space is
highly non-trivial and
very difficult in reality.
However, understanding the
ground state alone allows one to decide, for example, whether we deal with
a superfluid. This is our strategy here.

\section{Magnetic component of the plasma}

At high temperatures and for static quantities, all the non-perturbative physics is expected to be
described in terms of a three-dimensional theory \cite{pisarski}:
\begin{equation}\label{pisarski}
L~=~\frac{1}{g_3^2} \Big( \frac{1}{2} {\mathrm{Tr}} \, F_{ij}^2+  |D_i\Pi^a|^2+ V(\Pi^2)\Big)~~,
\end{equation}
where $\Pi^a$ is a scalar color field ($\omega_M=0$ component of the potential $A_0^a$).
As is mentioned above, the dimensional reduction implies simple scaling laws
for various quantities. In particular, if one defines magnetic monopoles
within the 3d Yang-Mills theory (which is a part of (\ref{pisarski}))
then the monopole density should scale as $g_3^6$
in terms of the dimensionally reduced coupling~\eq{g3}.
And, indeed, in the 3d Yang-Mills theory~\cite{grigorev}:
\begin{equation}\label{grigorev}
\rho_{3,{\mathrm{mon}}}~\approx ~10^{-7} g_3^6~~.
\end{equation}
According to~\eq{g3} the density is proportional to $T^3$ as would be also the case for massless particles
at temperature $T$. However, the density (\ref{grigorev})
is not given by the Boltzmann distribution in terms of the original temperature.
The temperature dependence arises because of the rescaling the fields of the original
4d theory. This trivial observation becomes crucial later.

As is mentioned in the Introduction, within the dual model of Yang-Mills theory
there exists an  absolutely different mechanism of reducing the non-perturbative physics
from four
to three dimensions~\cite{foot2}. It is related to the dynamics of strings (which are the basic objects
of the dual formulation, or Yang-Mills theories in the infrared).  To  put it shortly, instead of 2d magnetic surfaces
or strings percolating in 4d at $T=0$ one has at $T\geq T_c$ particles
percolating in 3d. Such a percolation can be adequately described
by 3d field theories. This conclusion, as is argued in \cite{atsushi},
arises  within  various
approaches, such as models of gauge/string dualities \cite{sasha}, effective field
theories, see in particular \cite{deforcrand}, or approaches based  on the lattice data
as referred to in \cite{atsushi}. For the sake of our presentation we will
reiterate the main points in the language of the lattice defects,
i.e. 2d surfaces (strings) or 1d trajectories (monopoles) mentioned above.

It is useful to start from the $T=0$ theory of confinement. Confinement is commonly believed to be due to condensation
of the magnetic degrees of freedom. Usually one understands by the magnetic
degrees  of freedom either Abelian monopoles or $Z_2$ vortices, for a review see,
e.g., \cite{greensite}. In fact  both
projections are manifestations of one and the same non-Abelian object.
Phenomenologically,
the monopole trajectories cover densely the vortices (2d surfaces) or, vice versa,
the vortices can be defined as minimal-area surfaces spanned on the monopole
trajectories, for a review see \cite{zakharov}
Both the vortices and monopoles percolate through the vacuum state, i.e.
form infinite clusters of the 2d surfaces or 1d trajectories. Important
properties of these clusters is that their total length, respectively area,
scale in physical units:
\begin{eqnarray}
L_{\mathrm{tot}}^{\mathrm{mon}}~ & \approx & ~ {\mathrm{const}} \cdot \Lambda_{QCD}^3V^{(4)}_{\mathrm{tot}}\,, \\
A_{\mathrm{tot}}^{\mathrm{vort}}~ & \approx & ~ {\mathrm{const}} \cdot \Lambda_{QCD}^2V^{(4)}_{\mathrm{tot}}\,,
\end{eqnarray}
where $V_{\mathrm{tot}}^{(4)}$ is the total 4d volume of the lattice.
The detailed picture for confinement does depend on whether one uses
monopoles or vortices. The existence of two alternative languages,
as we will see, is important within the context of this note.

What happens at $T\geq T_C$ is that the 4d percolation of the defects  is becoming a
3d percolation. In more detail and in the monopole language \cite{magncom,alessio}
the 4d percolating cluster disappears. Which corresponds to destroying
the condensate of the magnetically charged field by temperature.
Instead there appear
monopole trajectories which are wrapped around the periodic time direction.
One can argue that the wrapped trajectories in the Euclidean space correspond
to real particles in the Minkowski space \cite{magncom}.
The 3d density of the wrapped trajectories scales indeed in physical units \cite{alessio}:
\begin{equation}\label{wrapped}
\rho_{\mathrm{wr}}~\approx~T^3 f(T,\Lambda_{QCD})~,
\end{equation}
where
the function $f(T,\Lambda_{QCD})$ is slowly varying at high temperatures. It is crucial however that this function
does not depend on the lattice spacing, as it should be for physical objects.

Phenomenologically the geometrical picture simplifies actually further. First, already
at $T$ close (and larger than) $T_c$ the wrapping number is equal to $n_{\mathrm{wr}}= 1$
for practically all the wrapped monopole trajectories (while generically $n_{\mathrm{wr}}$ could be equal
to any integer). Moreover, the trajectories are rapidly becoming more and more static.
Roughly speaking, the approximation of static trajectories is not so bad beginning
with $T=T_c$~\cite{foot3}.

As is mentioned above, in the static limit one can consider a 3d picture.
In a 3d time slice the monopole trajectories become point-like excitations
which can be called instantons (in resemblance to but not in an exact correspondence
with the Polyakov model \cite{polyakov}).  The density (\ref{wrapped})
becomes the density of the instantons

Within the vortices, or string~\cite{foot4} approach the geometrical picture is similar, with the corresponding
change of dimensions. At $T>T_c$ the percolating vortices become preferably time oriented and,
moreover, simply static to a reasonable approximation. In the static approximation,
the 2d surfaces can be replaced by their 1d intersections with a given time-slice.
The 1d defects or trajectories correspond to particles, or fields in any number of dimensions.
Thus, the vortices reduce to a 3d field. While the time-direction dependence becomes trivial, the percolation
in the three spatial  dimensions persists. In field theoretic language this means that
the corresponding 3d scalar field, $\Sigma_M$ has a non-vanishing vacuum expectation value:
\begin{equation}\label{atsushi}
\langle \Sigma_M\rangle~\neq~0~,
\end{equation}
for further details see~\cite{atsushi}.

To summarize, the 3d physics sets in for non-perturbative effects
quite early, at temperatures, just above $T_c$
The reason seems to be understood rather within dual models
than within a field theoretic formulation.
In the region, say,
$$T_c~<T~<~2T_c$$
the parameters of the 3d field theory are to be treated phenomenologically. At larger temperatures
simple scaling laws like (\ref{grigorev}) become valid in many cases.

\section{Temperature dependence}

Imagine for a moment that the  3d magnetic component of plasma
is indeed a liquid, as is argued on the basis of the lattice data
\cite{magncom,shuryak}. Could it be a superconducting liquid?
At first sight, the answer is obviously ``no''. Indeed,  ordinary
superfluidity is destroyed at finite, but low temperature. But now
we are discussing temperatures of order $T_c~\sim~200$\,MeV.
However, why does  superfluidity, present at $T=0$, disappear at some $T_0$
despite of the fact that the two motions (superfluid and dissipative)
are independent? The reason \cite{lifshitz} could be called a kind of a ``nonrelativistic
unitarity condition''. The density of the bosons in the condensed mode $n_0$ diminishes
with temperature,
\begin{equation}
[n_0(0)-n_0(T)]/n_0(0)~\sim~T^2 \qquad (T\ll T_0)\,,
\end{equation}
and at $T=T_0$ the boson condensate gets evaporated, $n_0(T_0)=0$.

In the case of Yang-Mills theory and in the weak-coupling limit,
$g^2(T)\to 0$ the total energy/pressure
density starts with the Boltzmann's factors for non-interacting gluons.
One calculates then corrections in $g^2(T)$ and as  less and less uncertainty
is left in the perturbative sum the  phase space available for the non-perturbative
component (let it be superfluid) diminishes. However, as is mentioned above, the uncertainty
of the perturbative series does not go down as an inverse power of $T$~\cite{foot5} but
is proportional only to $g^6_3(T)$,  or to
$T^3 \ln (\Lambda_{QCD}/T)^{-3}$.
Thus, the r.h.s. of Eq.~(\ref{fit}) of characterizes the phase space of the component which
is actually  not controlled by temperature and is determined by the physics in the infrared even
at $T\to \infty$.

Thus, for the phase space factor associated with the magnetic component in Eq.~(\ref{independent})
we have
\begin{equation}
(phase~factor)_{\mathrm{magnetic}} ~\sim~\Big({1\over \ln T}\Big)^3~~,
\end{equation}
which implies an amusing possibility of having superfluidity even at $T\to \infty$
provided that the 3d field theory behind the magnetic component corresponds to
a superfluid~\cite{foot6}.

\section{Dynamics of the magnetic component}

The dynamics of the instantons (monopoles) in 3d Yang-Mills theories has been investigated
numerically in \cite{ishiguro}. In the high-temperature limit
this is our magnetic component as well. One assumes that monopoles can be treated as
Abelian objects with a partition function of a Coulomb gas:
\begin{equation}\label{polyakov}
Z~=~\sum_{N=0}^{\infty}{\zeta^N\over N!}\big[\prod_a\int d^3 x^{(a)}\big]
\exp\Big(-{g_m^2\over 2}\sum_{a\neq b} q_a q_b D_{ab}\Big)\,,
\end{equation}
where $D_{ab} \equiv D(x^{(a)}-x^{(b)})$ is the scalar particle propagator,
$q_{a,b}$ are the monopole charges in units of elementary magnetic charge $g_m$, $|q_a|=1$, and
$\zeta$ is the fugacity. The model (\ref{polyakov}) is known to induce
confinement of external electric charges~\cite{polyakov}.

It was found \cite{ishiguro} that the lattice data can be fit by the model (\ref{polyakov}).
In particular, one can define the Debye screening mass $m_D$ of the magnetic plasma described by
(\ref{polyakov}). However, it turns out that $\rho_{3,{\mathrm{mon}}}/m^3_D\approx 0.03$ where $\rho_{3,{\mathrm{mon}}}$
is the 3d monopole density. The latter observation is in contradiction with the mechanism
of the Debye screening. Another weak point of the model is that it replaces the original
non-Abelian action by its Abelian projection.

In view of the observation $\rho_{\mathrm{mon}}/m_D^3\ll 1$
one might be tempted to try an opposite limit and consider the system of the 3d monopoles
not as a plasma but rather as a Bose-particles  system with low density.
Then one of the possibility is the Bose condensation and, as a result, superfluidity
\cite{lifshitz}. The problem is tractable provided that the interaction region is small
compared to the volume occupied by a particle on average:
\begin{equation}
\rho_{\mathrm{mon}}a^3_{\mathrm{sc}}~\ll~1~~,
\end{equation}
where $a_{\mathrm{sc}}$ is the scattering length. Also, the interaction is to be repulsive, $a_{\mathrm{sc}}>0$.
Otherwise, the slow particles would attract each other and the condensation
of the original particles  is impossible.
In the Abelian projection, monopoles and anti-monopoles attract each other at short distances,
and the Bose condensation seems to be ruled out.

However, it is more consistent to view the ``monopoles'' and ``antimonopoles''
as non-Abelian objects detected through the Abelian projection, see, e.g., \cite{adriano}
and references therein. Then their interaction at short distances should be treated
phenomenologically, through the lattice studies. It was observed \cite{alessio}
that {\it both} monopole and monopole and monopole and anti-monopole repel each other
at short distances. In the language of the scattering lengths:
\begin{equation}
a_{\mathrm{mon-mon}}~>~0,~~ a_{\mathrm{mon-antimon}}~>~0~,
\end{equation}
and there is no contradiction, at this level, with the idea of the Bose condensation.
A reservation is that monopole and  antimonopole still attract each other at
``intermediate distances'', while the monopole-monopole interaction is repulsive
at all distances.  The attraction, however, is not strong enough
to bind the particles and in this sense can be neglected.

Moreover, it turned also possible \cite{alessio} to extract the interaction potentials.  In case of the monopole-monopole interaction the estimate is:
\begin{equation}
V_{\mathrm{mon-mon}}(r)~\sim~{1\over r}\exp (-r/\lambda)~~,~~\lambda~\approx~0.1~{\mathrm{fm}}~.
\end{equation}
Thus, one can estimate the interaction region as
\begin{equation}\label{region} V_{\mathrm{int}}\sim (0.2~{\mathrm{fm}})^3~~.
\end{equation}
Whether the interaction region (\ref{region}) is large or small depends
on the density of the monopoles which is also provided by the lattice measurements
\cite{alessio}.

As is mentioned above at high temperatures we expect that all the temperature dependencies
are trivial. Namely, after rescaling fields and distances the 3d theory does not depend
on the temperature at all. In  other words, all the observables should be proportional to
the corresponding powers of $g^2(T)\cdot T$. This should be true also for the
interaction region (\ref{region})
However, numerically there is no much evidence for that. It is more
appropriate to say that we are dealing with estimates rather than with exact numbers.
The density of the monopoles, on the other hand, is measured with good accuracy.
In particular,
$$\rho_{\mathrm{mon}}~<~5~{\mathrm{fm}}^3, ~~~~T_c~<T~<~2T_c~.$$
Thus, in this temperature range the approximation $V_{\mathrm{int}}\cdot \rho_{\mathrm{mon}} ~\ll~1$ seems
granted. For higher temperatures the issue is more subtle and we postpone a detailed discussion
of the numerics. The general impression is that the density is still low enough in
the appropriate units.

Now we come to the following question. Phenomenologically, we have
two descriptions of the medium of 3d excitations. If we start from the 4d monopoles
at $T=0$
 then they become a 3d gas of instantons at $T>T_c$. On other hand, if we start
from magnetic strings then they turn into a 3d magnetically charged scalar field
with non-trivial vacuum expectations value (\ref{atsushi}).
Both description are obtained in Abelian projections and
in this sense  both oversimplify the actual non-Abelian picture.
However, if we are looking for a possible match of the lattice phenomenology to
the theory of superfluidity, the first impression is that the pictures are mutually
excluding each other and only one of them has chances to be correct, if any.

The good news is that both descriptions can correspond to superfluidity and the two pictures
are just dual to each other, see, e.g., \cite{schakel}.
This duality is well known in the theory of superfluidity. One starts
with the Hamiltonian for heavy particles:
\begin{equation}\label{heavy}
\hat{H}~=~{p^2\over 2m}\hat{a}^{+}_{\bf p}\hat{a}_{\bf p}+{U_0\over 2V}
\Sigma\hat{a}^{+}_{\bf p_1^{'}}\hat{a}^{+}_{\bf p_2^{'}}\hat{a}_{\bf p_1}\hat{a}_{\bf p_2}~,
\end{equation}
 where $\hat{a}, \hat{a}^{+}$ are annihilation and production operators.
 One performs then the Bogolyubov transformation,
\begin{equation}
\hat{a}_{\bf p}~=~u_{\bf p}\hat{b}_{\bf p}+v_{\bf p}\hat{b}^{+}_{-\bf p},~~\hat{a}_{\bf p}^{+}~=~
u_{\bf p}\hat{b}^{+}_{\bf p}+v_{\bf p}\hat{b}_{-\bf p}~~,
\end{equation}
where $u_{\bf p},v_{\bf p}$ are coefficients, to diagonalize the Hamiltonian.
In terms of the new field (associated with the operators $\hat{b},\hat{b}^{+}$)
the spectrum starts with linear, or phonon term:
$$\epsilon({p})~\approx~up~~,$$
where $u$ is the speed of sound.
The new field, associated with the operators $\hat{b},\hat{b}^{+}$
has vacuum expectation value
and we can identify this field phenomenologically with the field $\Sigma_M$, see Eq. (\ref{atsushi}).

Thus, the existence of the two descriptions of the ground state, in terms of
the gas of monopoles/instantons and in terms of an infinite cluster
of trajectories, or vacuum expectation value (\ref{atsushi})
is in fact a strong argument in favor of the superfluidity of the
magnetic component.

Note that the linear spectrum $\epsilon (p)=u\cdot p$
corresponds to a massless field in ``relativistic 3d language''. This masslessness can be traced back to the
magnetic $U(1)$ symmetry of Hamiltonian (\ref{heavy}) in terms of heavy particles.
As is mentioned above the actual `monopole'' and ``antimonopole''
do not interact at short distances as a particle and antiparticle
(the terminology used is rooted in the Abelian projection but the
actual non-Abelian dynamics is different).
As a result, the would-be massless Goldstone boson does not show up
in the spectrum of the 4d theory.

Because confinement in the spatial directions is due to breaking of
magnetic $U(1)$ to magnetic $\mathbb{Z}_2$
a phenomenological 3d model which seems to be more
appropriate to describe the condensation (\ref{atsushi})
is the 't~Hooft model~\cite{thooft}:
\begin{eqnarray*}\label{model}
  \mathcal{L} &=& \partial_\mu \varphi \partial_\mu\varphi^*-M^2\varphi\varphi^*-\lambda(\varphi\varphi^*)^2
  +\frac{\zeta}{2}\left((\varphi)^2+(\varphi^*)^2\right)
\label{thooft}
\end{eqnarray*}
This model has the  magnetic U(1) broken to $\mathbb{Z}_2$
by the $\zeta$-term and
incorporates 3d confinement (area law for the spatial Wilson loop).
There is no Goldstone particle. The model (\ref{model}) might
describe condensation of the field $\Sigma_M$.

\section{Conclusions}

It appears that the magnetic component of the Yang-Mills plasma could
provide an independent  component to the fluidity of the plasma. It is most remarkable
that the properties of this component, especially its viscosity, are
in a way independent of the temperature. The temperature does determine
the phase factor which controls the contribution of this component of plasma to the
total viscosity~\eq{independent} but not the partial viscosity itself. The reason is that
the magnetic component is directly related to the infrared divergences known
since long time in high-temperature field theory. As a result the density
of, say, monopoles is not given by a Bose distribution corresponding to the
overall temperature and certain mass of the monopole. Instead, it is proportional to $(g^2(T)\cdot T)^3$
at high temperatures. To adjust phenomenology to this prediction of the theory
one can introduce a corresponding chemical potential   \cite{magncom}
but this is just another demonstration that the density of the magnetic degrees
of freedom is not determined by the standard  high-temperature dynamics.

The properties of the magnetic component are determined
by a 3d field theory. At very high temperature it should be
the standard dimensionally reduced Yang-Mills theory.
At intermediate temperatures, parameters of the 3d theory
can be fitted phenomenologically. In particular, the magnetic component
could be superfluid. To clarify whether such a possibility realizes we
invoke lattice data. Because the data are obtained in Euclidean space
they refer in fact to the ground state. Phenomenologically superfluidity
of the magnetic component seems plausible although further data are required
to make the evaluation more reliable.

The same lattice data seem to fit known Abelian models of three-dimensional confinement
as far as long-distance interaction of the constituents is concerned.
At short distances, the actual non-Abelian nature of the magnetic degrees of
freedom is manifested and turns crucial for the self-consistency of the models.
Thus, there is a perspective that the same magnetic component of the Yang-Mills
plasma could explain both the 3d confinement and low viscosity of the plasma.

\acknowledgments
We are grateful to M.~D'Elia, A.S.~Gorsky, F.V.~Gubarev, A.~Nakamura and A.~Niemi
for useful discussions. The paper was worked out during the visits
of V.I.Z. to the Universities of Gent, Belgium and Tours, France.
V.I.Z is thankful for the hospitality extended to him during these visits.

This work was supported by the STINT Institutional grant IG2004-2 025,
by the RFBR 08-02-00661-a, DFG-RFBR 436 RUS, by a grant for scientific
schools NSh-679.2008.2, by the Federal Program of
the Russian Ministry of Industry, Science and Technology
No. 40.052.1.1.1112 and by the Russian Federal Agency
for Nuclear Power.


\begin{thebibliography}{99}
\bibitem{foot1}
The lattice data refer mostly to pure Yang-Mills theories and
we will concentrate in fact on the $SU(2)$ case assuming that inclusion of dynamical quarks does not
change the basic features of the plasma.
\bibitem{foot2}
And, probably, this is the explanation for the ``precocious'' validity
of the asymptotic, $T\to\infty$ behavior of non-perturbative observables.
\bibitem{foot3}
We describe the data for the $SU(2)$ case. For larger
number of colors one expects that these features become only more conspicuous.
\bibitem{foot5}
This would be an analogy to the nonrelativistic case.
Also at $T=0$ the uncertainty of the perturbative series is power-like.
\bibitem{foot6}
Having superfluidity does not mean $\eta_{magnetic}\equiv 0$
since there are quantum corrections, for references and discission see, e.g., \cite{teaney}.
\end{thebibliography}
\end{document}